\documentclass[referee]{raa}

\usepackage{graphicx,times}        
\usepackage{natbib} 
\usepackage{amssymb,amsmath}
\usepackage{caption}
\usepackage{subcaption}
\usepackage[pagebackref=true]{hyperref}

\begin{document}

      \title{Variation of small scale magnetic fields over a century using Ca-K images as proxy}
	\volnopage{Vol.0 (20xx) No.0, 000--000}      
  	\setcounter{page}{1}          

  	 \author{Jagdev Singh\inst{1}, Muthu Priyal\inst{1}, B. Ravindra\inst{1}, Luca Bertello\inst{2}, Alexei Pevtsov\inst{2}}

		\institute{Indian Institute of Astrophysics, II Block, Koramangala, Bengaluru – 560 034, India; {\it jsingh@iiap.res.in}\\
    		 \and 
    		         National Solar Observatory, 3365 Discovery Drive, Third Floor, Boulder, CO 80303, USA\\
\vs\no
   {\small Received 20xx month day; accepted 20xx month day}}

\abstract{
A combined uniform and long-time series of Ca-K images from the Kodaikanal Observatory (KO), Mount Wilson Observatory (MWO), and Mauna Loa Solar Observatory (MLSO) were used to identify and study the Ca-K small-scale features and their solar cycle variations over a century.  The small scale features are classified into three distinct categories$:$ enhanced network (EN), active network (AN), and quiet network (QN). All these features show that their areas vary according to the 11-year solar cycle. The relative amplitude of the Ca-K network variations agree with that of the sunspot cycle. The total area of these small-scale features varies from about 5\% during the minimum phase of the solar cycle to about 20\% during its maximum phase. Considering the average intensity and the amplitude of their area variations, we find that the total contribution of EN, AN and QN to the irradiance variation of the Sun is about 3\%. 
  }

\authorrunning{Singh et al.}
\titlerunning{\textit{Ca-K small scale} features}
\maketitle

\section{Introduction}
The variable magnetic field of the Sun causes processes that influence the heliosphere and the Earth’s upper atmosphere \citep{bertello2020}. The main cause for this variability is the complex dynamo process, believed to be running at the bottom boundary layer of the solar convection zone, known as the tachocline \citep{gilman2000, weiss2000}. The operation of the dynamo process leads to the formation of sunspots and other related activity on the solar surface \citep{choudhuri1995}. The occurrence and characteristics of the sunspot cycle have been studied by several researchers \citep{howard1984, howard1992, howard1993, howard1999, howard2000, sivaraman2003} using long data series spread over several decades.

\par  Apart from the large-scale activity in the form of sunspots and related plage regions visible on the solar surface, the quiet Sun shows a reticular pattern of intense kilo-gauss fields known as the magnetic network and small-scale flux concentrations in the areas in between them \citep{rubio2019}. A significant fraction of the photospheric magnetic flux resides in the quiet Sun in the form of a magnetic network \citep{gosi2014}. According to \cite{jin2011}, the quiet Sun magnetic field is comparable to the total flux of active regions during the maximum phase of the solar cycle. The emergence and cancellation of the quiet Sun magnetic field couple the different atmospheric layers and may contribute effectively to heating the chromosphere and the corona.

 Regular observations of the Sun's magnetic field began in the late 1960s at the Crimean Astrophysical Observatory \citep{kotov1998}, the Mount Wilson Observatory \citep[MWO]{livingston1976}), the Wilcox Solar Observatory \citep[WSO;][]{scherrer1977}, and a few other observatories. As time progressed, techniques and accuracy in measuring the small-scale magnetic field have improved \citep{plotnikov2021}. \cite{schrijver1998} computed the emergence rate of small bipolar regions and argued that the frequent magnetic reconnections associated with these regions would heat the solar atmosphere. Thus, knowledge of long-term variation in the small-scale magnetic field is vital to understanding the solar atmosphere's heating.
 
Magnetic elements appear bright in chromospheric images as indicated in Ca-K \& H filtergrams in comparison to photospheric magnetograms \citep{babcock1955, stepanov1959, leighton1959, leighton1964}. \cite{sivaraman1982} showed a strong correlation between the Ca-K bright points and small-scale magnetic fields. The plage and networks observed in the Ca-K line can be used as a proxy to study the variations in magnetic fields as there is a strong correlation between Ca-K intensity and magnetic fields \citep{skumanich1975, schrijver1989, ortiz2005}. Long-term variations in large-scale activity using Ca-K plage areas have been reported in several studies \citep{foukal1996, tlatov2009, ermolli2009, foukal2009, chatzistergos2019}. The variations in the Ca-K index (including plages and small-scale network regions) have been studied by \cite{bertello2010}. \cite{pevtsov2016}, using the polarity of sunspot magnetic field data and Ca-K intensity, showed that a long-time series of pseudo-magnetograms can be constructed by making use of Ca-k data. 

Using 1400 digitized Ca-K spectroheliograms obtained at the Sacramento Peak Observatory between 1980 and 1996,  \cite{worden1998} classified the plage regions, enhanced network (EN), active network (AN), and quiet Sun by defining an intensity threshold value for each of these features.  \cite{priyal2017, priyal2019} attempted to study the variations in the plage and network areas using the digitized Ca-K spectroheliograms obtained at Kodaikanal Observatory (KO) during the period 1905 -- 2007. The averaging of the measured parameters over several months was required because of the large scatter in their values due to different contrast of the obtained images. The monthly averaged Ca-K parameters were used by several investigators to study the long-term variations \citep{priyal2019, pooja2021}.

\par \cite{bertello2020} rescaled the intensity of Ca-K  MWO images such that the FWHM of the intensity distribution of each image was 0.08. They selected 14525 images of uniform quality to study the variation in the Ca-K index and chromosphere's differential rotation. \cite{singh2021} developed an equal contrast technique (ECT) that makes the intensity distribution of the background chromosphere similar in all images. First,  images are subjected to density-to-intensity conversion using step wedge calibration \citep{priyal2014} and then corrected for the large-scale variations in intensity due to limb darkening and instrumental vignetting. The maximum intensity of the image is normalized to unity. Then, the contrast of all the images was made uniform using the ECT methodology, which involves rescaling the intensities of the image, such that the FWHM of the intensity distribution lies between 0.10 – 0.11 \citep{singh2021}.

\cite{singh2022} showed that data of different observatories such as KO, Mount Wilson Observatory (MWO), and Mauna Loa Solar Observatory (MLSO) could be merged after using the ECT methodology to reduce time and data gap and making long series of consistent quality data.

 Further, \cite{singh2022} studied the variations in the large-scale magnetic field using Ca-K plages as a proxy. In this paper, we study the variations in small-scale features such as enhanced network (EN), active network (AN), and quiet network (QN) areas with time. It is generally believed that EN represents small regions of decaying plages, AN represents small bipolar regions, and QN defines the boundaries of the network. In section 2, we summarize the data and its analysis. Section 3 describes the results of data analysis, and in section 4, we discuss the implications of the results.  

\section{Data Analysis}

Here, we investigate five sets of Ca-K line images from three observatories, KO, MWO, and MLSO. The images are Ca-K spectroheliograms obtained at KO from 1905 to 2007 and MWO from 1915 to 1985. We have also considered Ca-K filtergrams taken at KO from 1997 to 2013 and MLSO with PSPT from 1998 to 2015. The details of data and analysis may be seen in \citet{singh2022}. 

We have used the data from several observatories to study the long-term variations of the chromospheric features. \cite{bertello2020} have rescaled the intensity of the images such that the FWHM of the intensity distribution becomes 0.08. Using this methodology, they could study about 40\% of the images available at MWO. We have rescaled the intensity of images using a different procedure, named Equal Contrast Technique (ECT) such that the FWHM of the intensity distribution lies between 0.10 - 0.11 as the median value of the FWHM considering the data of MWO and KO is $\sim$0.1. Using the ECT procedure, we could study all the available images.

\par
\begin{figure}[ht!]
\centerline{\includegraphics[width=0.8\textwidth]{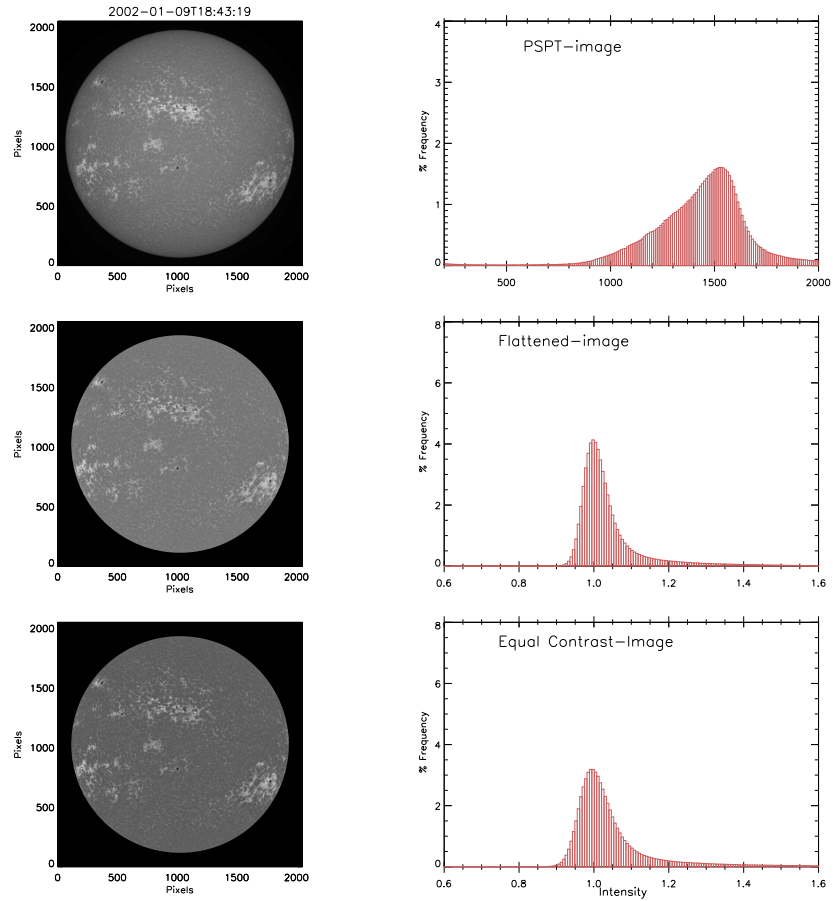}}
\caption{ Top two panels of the figure show the Ca-K image taken on December 24, 2003 with PSPT at MLSO and its intensity distribution. Middle two panels indicate the flattened image using the code developed for KO data and intensity distribution. Bottom two panels show the image and its intensity distribution after using the ECT methodology developed for KO data.}
\label{fig:1}
\end{figure}

 The top two panels of Figure \ref{fig:1} show a digitized image and its intensity distribution taken on December 24, 2003, with PSPT at MLSO. The middle two panels show the corrected image for the instrumental vignetting, (using the code developed for KO data) and its intensity distribution. The intensity distribution indicates relatively narrower FWHM, a signature of low contrast images. The bottom two panels show the image after adjusting the contrast such that the FWHM of the intensity distribution lies between 0.10 – 0.11.

The plages were identified as regions with normalized intensity $>$ 1.30 and area $>$ 0.2 arcmin$^{2}$, EN with intensity $>$ 1.30 and area $>$ 4 pixels (3 arcsec$^{2}$). The AN is defined as regions with intensity between 1.2 and 1.3 and QN is between 1.1 and 1.2 with an area $>$ 4 pixels (3 arcsec$^{2}$) \citep{singh2021}. The areas were determined in full-disk images in pixels in the plane of the sky. A lower limit of 4 pixels has been used to avoid the random contribution of single pixels. The regions with intensity $<$ 1.1 represent background (quiet) chromospheres. Here, we investigate the variation in small-scale features EN, AN, and QN as a function of time and their correlation with sunspot numbers from 1905 to 2015 by combining all the data sets.

\begin{figure}[ht!]
\centerline{\includegraphics[width=0.8\textwidth]{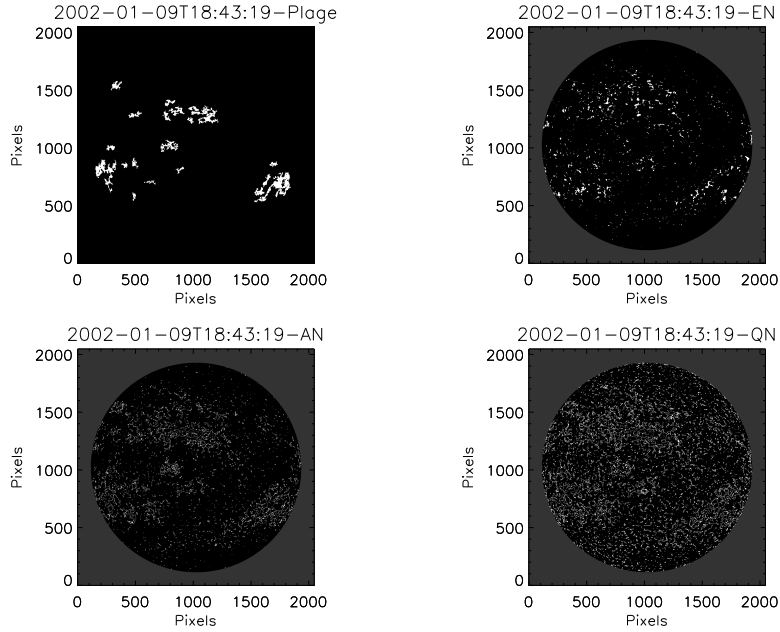}}
\caption{ Binary images of the identified plages, EN, AN, and QN for the Ca-K image obtained on December 24, 2003 using PSPT at MLSO.}
\label{fig:2}
\end{figure}

Figure \ref{fig:2} shows the detected plage, EN, AN, and QN areas in white color in the binary format, using the intensity as mentioned earlier and area threshold values from a typical image obtained at PSPT on December 24, 2003. The top-left panel of the figure shows the large-scale plage area, generally seen in the active regions on the sun. The top-right panel of these figures shows the EN regions in white. The EN regions in the vicinity of plages imply that these are fragments of the decaying plages. The bottom two panels of the figure show the AN, and QN area. AN regions are mostly concentrated in the equatorial (active) regions, whereas QN regions are seen all over the sun representing the quiet network.

\section{Results}

First, we compare the measured values of EN, AN, and QN areas before and after the ECT application between KO and MWO spectroheliograms  to find the effectiveness of ECT in detecting small-scale features and then to study the variations of these parameters with time. Afterward, we combine the ECT corrected data of KO, MWO, and PSPT and examine their correlation with the long-standing record of sunspot data. 

\subsection{Comparison of network parameters for KO and MWO data before and after the ECT application}

\begin{figure}[ht!]
\centerline{\includegraphics[width=0.8\textwidth]{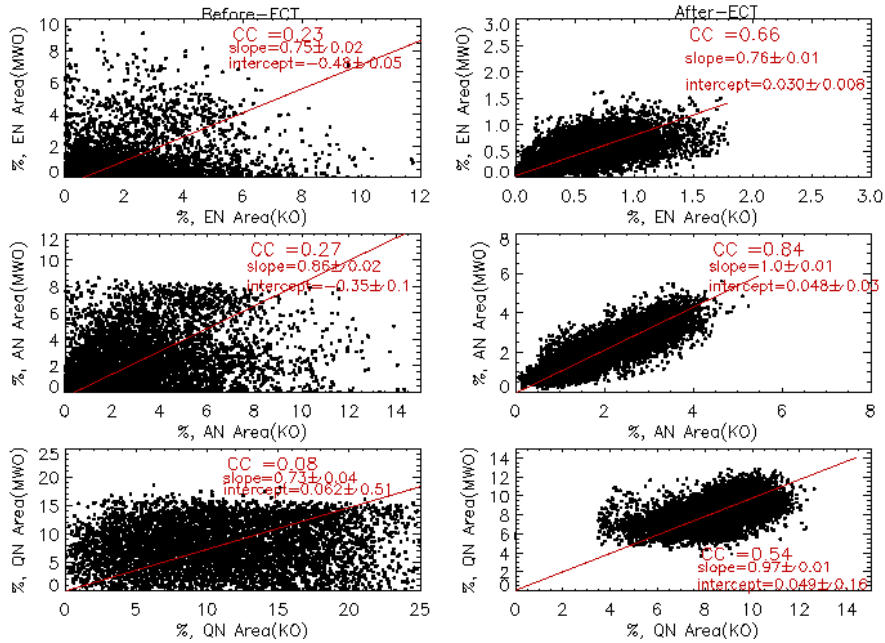}}
\caption{The left and right panels in the top row of the figure show the EN area in percentage of solar disk detected from KO spectroheliograms versus that from MWO data before and after the ECT application, respectively, for the period 1915 – 85. Two panels in the middle row show the scatter plots for the \% AN area before and after the ECT application. Two panels in the bottom row show the same for \% QN area. It may be noted that plots after the ECT application are on magnified scale (about factor of 1.8) as compared to before the ECT.}
\label{fig:3}
\end{figure}

The top two panels in Figure \ref{fig:3} show the scatter plot of EN area in the percentage of solar disk for the KO and MWO data before and after the ECT application. The values of correlation coefficients, intercept, and slope of the linear fit (red line) are written in each panel. The values of correlation coefficients 0.23 and 0.66 for the EN area before and after the ECT application indicate a significant improvement in the correlation between the two data sets after the ECT. The two panels in the middle row of Figure \ref{fig:3} show the scatter plots of the AN area for the KO and MWO data before and after the ECT application. The values of correlation coefficients (0.27 before ECT and 0.84 after ECT) indicate a large improvement in the correlation after the ECT methodology. Similarly, scatter plots of the QN area for KO and MWO in the bottom row of the figure indicate much improvement in the correlation between the two data sets after the ECT application. The values of intercept (nearly zero) and the linear regression slope (0.76, 1.0, and 0.97 for EN, AN, and QN, respectively) after the ECT application imply that the detected values of these Ca-K parameters agree with each other. The comparison of KO and PSPT filtergram data shows similar results. 

\subsubsection{Variation of EN, AN, and QN areas with time for KO and MWO data}

\begin{figure}[ht!]
\centerline{\includegraphics[width=0.8\textwidth]{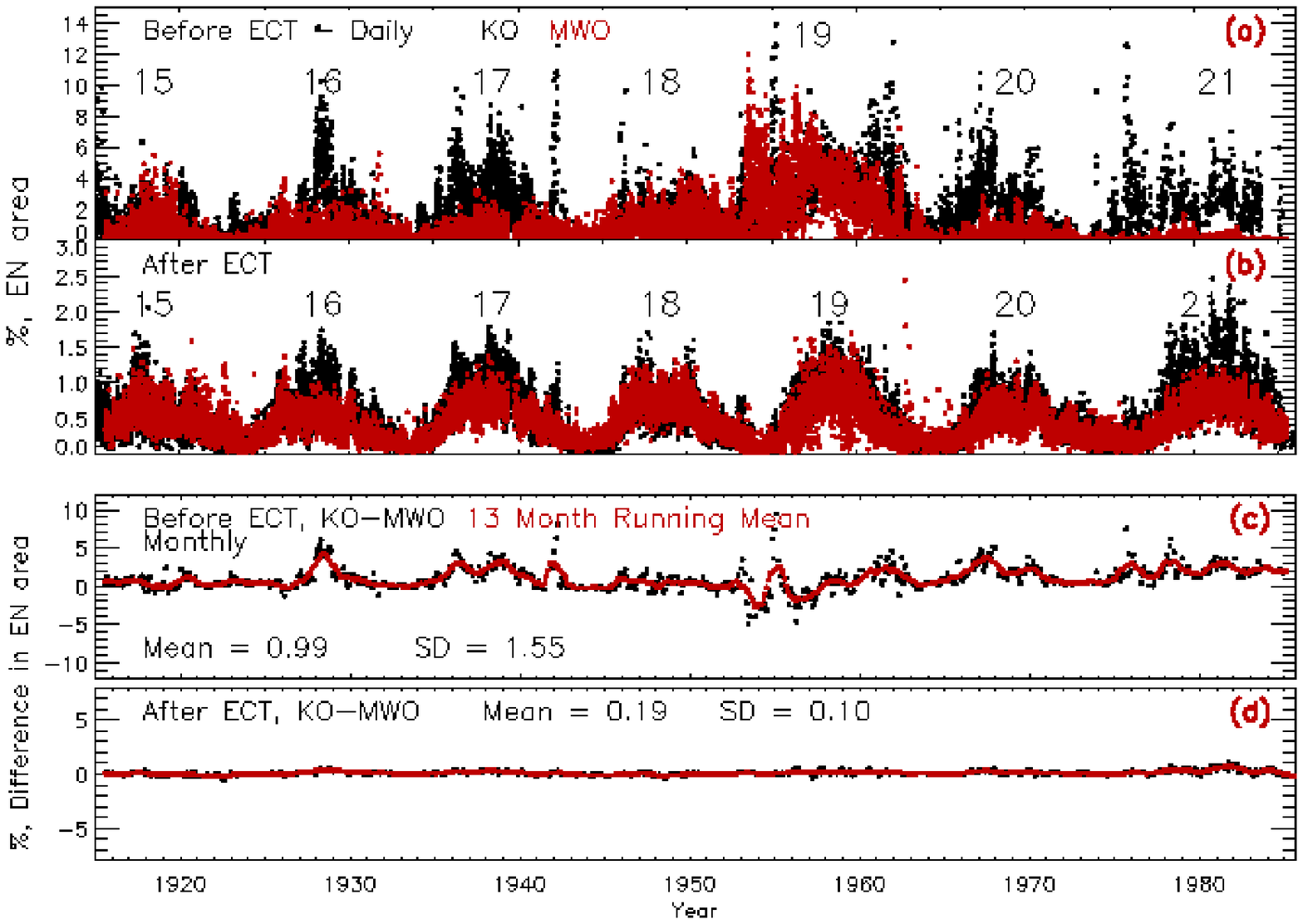}}
\caption{Panel (a) of the figure shows the percentage of EN area as a function of time for the period 1915 -- 1985, common period of observations at KO (black dots) and MWO (red dots) before the ECT application on a daily basis. Panel (b) indicates the percentage of EN area after the ECT application to the images. The solar cycle numbers are indicated in the panels. Panels (c) and (d) show a difference (KO – MWO) in percentage of EN area (black dots) on a monthly average and 13-month running average (red dots) basis before and after the ECT application, respectively. The mean and SD values are written in the bottom two panels.}
\label{fig:4}
\end{figure}

Panel (a) of Figure \ref{fig:4}  shows the percentage of measured EN area for the KO (black dots) and MWO (red dots) data before the ECT application. But after the ECT application, the maximum amplitude of the percentage of EN area decreases from $\approx$10 \% (before the ECT) to $\approx$2\%, as seen in panel (b). Panel (c) of this figure shows the difference (KO -- MWO) in the detected values of \% EN area on monthly (black dots) and 13-month running average (red dots) before the ECT application. The values show that the differences of \% EN area vary between – 4\% to + 4\%, even after averaging the data over a long period. Panel (d) shows that the difference in (KO -MWO) \% EN area after the ECT application on a monthly average basis varies between -0.5\% to +0.5\%. A small difference is considering the nature of data obtained with two instruments with different specifications and weather conditions. The mean value of the difference is 0.19\% with an SD value of 0.10\%. This suggests that the two data sets can be combined to decrease the gaps in the long-time series of Ca-K images.

\begin{figure}[ht!]
\centerline{\includegraphics[width=0.8\textwidth]{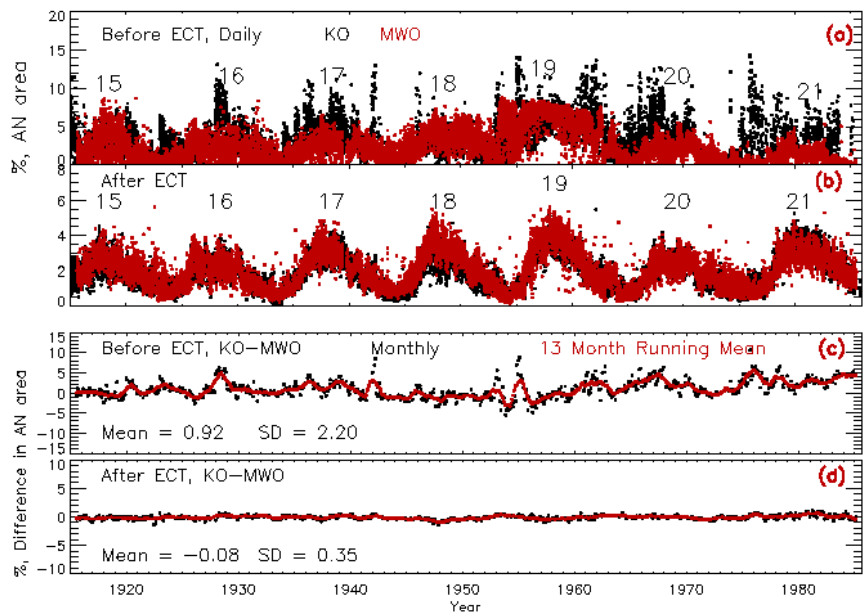}}
\caption{Same as that of Figure \ref{fig:4} but for AN area.}
\label{fig:5}
\end{figure}

\begin{figure}[ht!]
\centerline{\includegraphics[width=0.8\textwidth]{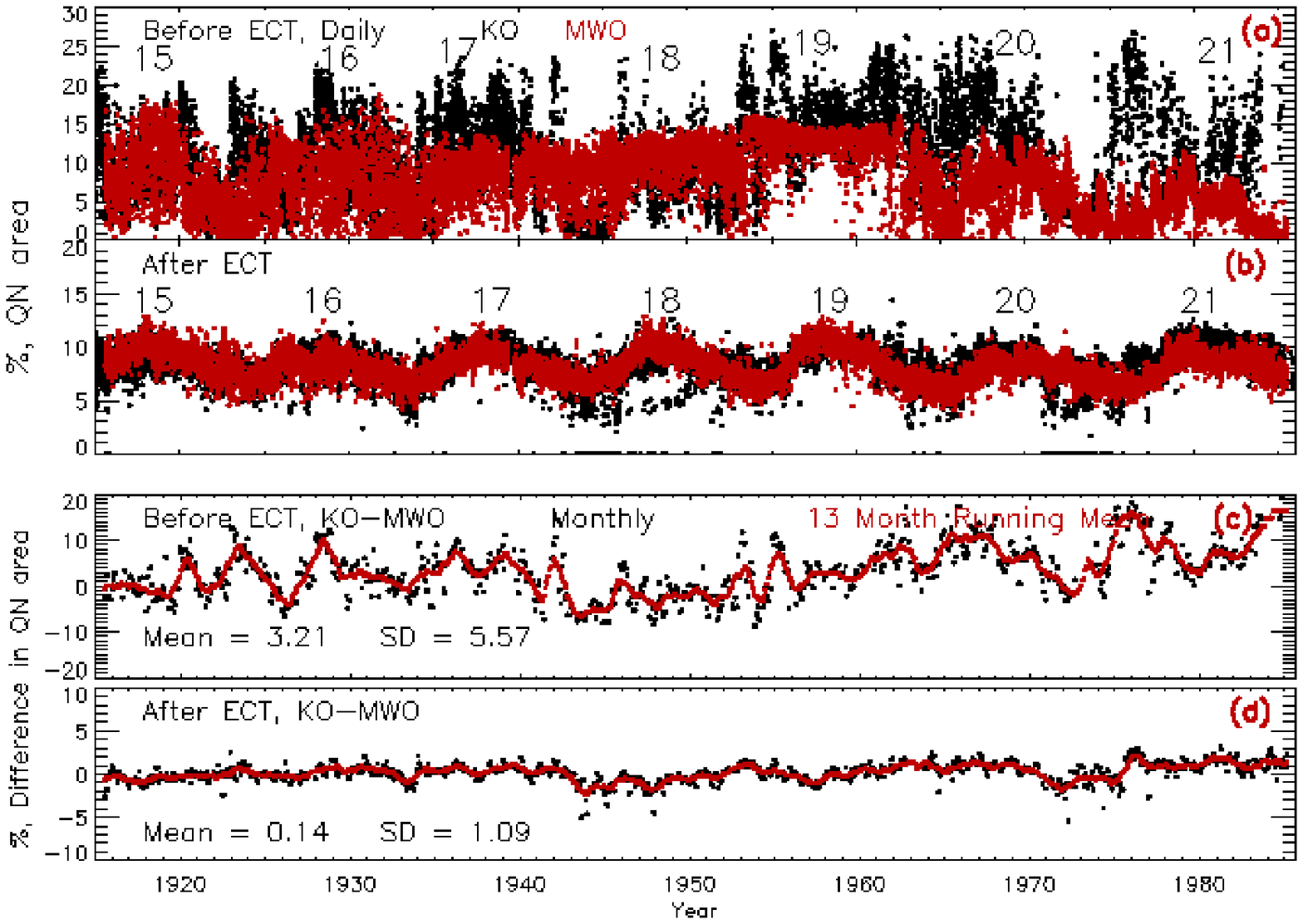}}
\caption{Same as that of Figure \ref{fig:4} but for QN area.}
\label{fig:6}
\end{figure}

Panel (a) of Figure \ref{fig:5} shows that detected values of \% AN area for KO and MWO data have significant differences before the ECT application. Panel (b) indicates that after applying the ECT, the detected values of the percentage of AN area for KO and MWO images agree with each other, even in daily data. The plot in panel (c) shows that the difference of (KO – MWO) of AN area on a monthly average is $\approx$ 4\% and increases after 1960. After the ECT application, the mean difference in KO and MWO values is $<$ 1\%, as seen in panel (d). Similarly, Figure \ref{fig:6} for the QN area shows a significant improvement in the agreement between the two data sets after the ECT application. The plot of the percentage of QN area as a function of time in panel (a) does not show significant solar cycle variations before the ECT application. But these become apparent after the ECT application, as seen in panel (b).

\subsection{Correlation between  Ca-K network parameters and sunspot numbers}

\begin{figure}[ht!]
\centerline{\includegraphics[width=0.8\textwidth]{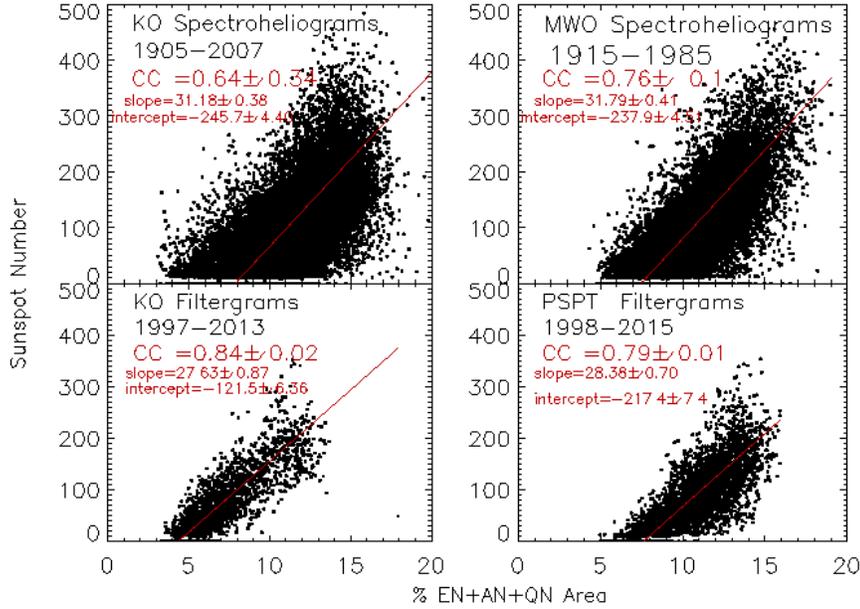}}
\caption{The left panel in top row shows the plot of percentage of small scale features (EN+AN+QN) area, hereafter called ``active area'' detected from KO spectroheliograms versus sunspot number on daily basis along with the linear fit (red line) to the data. The right panel shows the scatter plot between the percentage of EN area and sunspot number for MWO spectroheliogram. The bottom two panels show the scatter plot for KO and PSPT (MLSO) filtergrams. The values of correlation coefficient, slope and intercept of the linear regression are shown in respective panels.}
\label{fig:7}
\end{figure}

We compare the measured percentage of small-scale active (EN+AN+QN, hereafter called ``active area'') area after the ECT application with the sunspot number to determine the relationship between the two. The upper-row left and right panels of Figure \ref{fig:7} show a scatter plot of the daily active area expressed in percentage versus sunspot number (WDC-SILSO, Royal Observatory of Belgium, Brussels, https$:$//wwwbis.sidc.be/silso/) for the KO and MWO spectroheliograms, respectively. In the plot, the red line shows the linear regression. We see that the fraction of solar disk covered by active area pixels (EN+AN+QN) varies between about four and 15\%, of which the QN is the most significant contributor (Figure~\ref{fig:6}).

We have tried higher degree polynomial fits to these data sets but found that linear fit is best among these. It appears that the relationship between these two parameters is linear up to sunspot number $<$ 250, but after sunspot number $>$ 250, the percentage of the area of small-scale features remains the same or increases by a small percentage. The values of correlation coefficient 0.64 (KO) and 0.76 (MWO) indicate a confidence level of $>$ 95\%. Nearly the same values of slope and intercept for the KO and MWO data imply that the detected combined \% area of small-scale features agree with each other very well. We have performed a similar analysis using sunspot areas and found comparable results.

The bottom two panels of the figure indicate that the detected percentage of the area of small-scale features from the PSPT images is more than that from KO filtergrams by about 30\%. The reason for this may be the passband of the Ca-K filter, the spatial resolution of images, and the photometric accuracy of the detector.

\subsection{Solar cycle variations of Ca-K network area}

\begin{figure}[ht!]
\centerline{\includegraphics[width=0.8\textwidth]{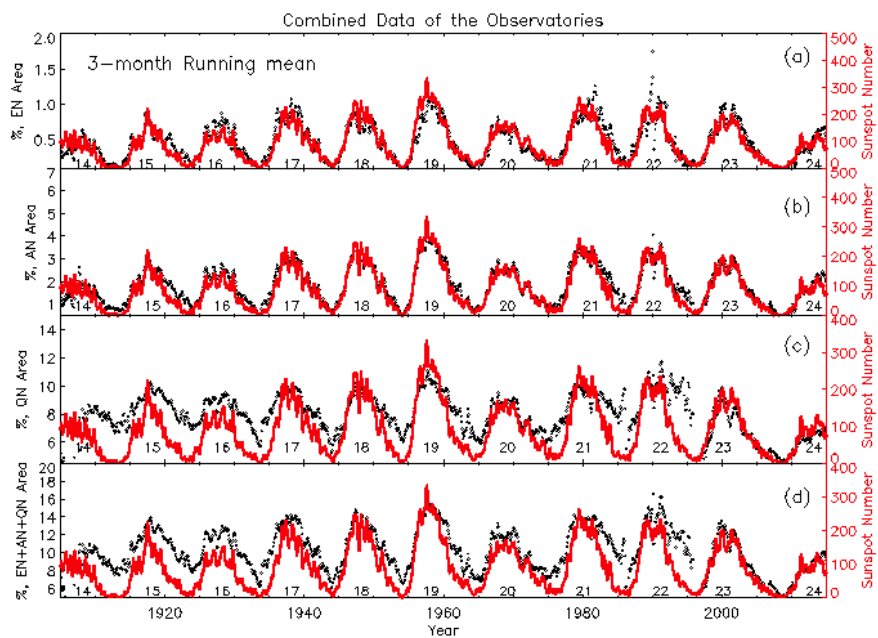}}
\caption{Panel (a) of the figure shows the percentage of EN area (black dots) as a function of time for the period 1905 -- 2015 on a monthly basis with running average over 3-months for merged data of KO, MWO and MLSO after the ECT application. The red curve shows the sunspot number for comparison. Panels (b) and (c) show the same for the percentage of AN and the percentage of QN area respectively. Panel (d) indicates the total percentage  of small scale features (EN + AN + QN) and the sunspot number in black and red, respectively. The solar cycle numbers are indicated in the panels.}
\label{fig:8}
\end{figure}

We have shown that the percentage area of solar disk occupied by small-scale features detected from the KO and MWO spectroheliograms agree with each other after the ECT application. To study the correlation of small-scale features over 1905 -- 2015 with reliable activity indexes such as sunspots,  we have combined the respective detected parameters of Ca-K small-scale features from spectroheliograms and filtergrams obtained at KO, MWO, and MLSO (PSPT) on a daily basis. Then we computed monthly averaged data with a running mean over 3-months for the Ca-K features and the sunspot numbers. Panel (a) of Figure \ref{fig:8} shows the percentage of EN area monthly (black dots), and the red curve indicates the monthly mean of sunspot number for comparison. The variations in the relative amplitude of both these parameters for solar cycles 14 - 24 agree with a minor difference.

The difference during cycle 21 is due to the quality of the KO data \citep{singh1988, priyal2019}. The \% EN area becomes nearly zero during the minimum phase of the solar cycle, and amplitude during the maximum phase varies between 0.6 to 1.1\% of the area of the visible disk when considered on a 3-monthly average basis. The plot of \% AN area and sunspot number (panel b) indicates that even during the minimum phase of the solar cycle, the AN area occupies about 0.5\% area of the visible disk, and the amplitude of the \% AN area at maximum phase varies between 2\% (solar cycle 14) and 4\% (solar cycle 19, the strongest cycle of 20th century). The variation in the relative amplitude of the percentage of AN area agrees well with those of sunspot numbers for all the cycles studied. Panel (c) indicates that \% QN area varies between 4 – 7 \% during the minimum of the solar cycle and 9 - 11 \% during the maximum phase during the 20th century. The plot indicates the area occupied by small-scale features is 5 – 8\% during the minimum and 10 – 16 \% during the maximum phase of solar cycles when averaged over three months.

\begin{table}[ht!]
\begin{center}
\caption{The values of FWHM of the intensity distribution before the ECT application, quiet network area for the KO and MWO data before and after the ECT application to the images are listed. }
\label{tab:1}
\vspace*{5mm}
\scalebox{0.85}{%

\begin{tabular}{ |c|c|c|c|c|c|c| }
 \hline
\textbf{Observation Date}  & \multicolumn{2}{c|}{\textbf{FWHM before ECT}}  & \multicolumn{2}{c|}{\textbf{Network area before ECT}} & \multicolumn{2}{c|}{\textbf{Network area after ECT}}\\

 \hline
   & \textbf{KO}  &	\textbf{MWO} &	\textbf{KO} &	\textbf{MWO} & \textbf{KO}  & \textbf{MWO} \\ 
 \hline
 1956-11-08 & 0.069 & 0.304 & 9.57 & 28.13 & 13.46 & 16.28 \\
\hline 
1956-11-25&0.056&0.302&6.53&26.77&13.68&15.88\\
\hline
1956-12-15&0.074&0.305&10.26&29.38&14.24&13.26\\
\hline 
1956-12-16&0.068&0.303&7.97&29.00&12.47&14.52\\
\hline 
1956-01-04&0.067&0.260&7.95&27.74&12.80&12.04\\
\hline 
1956-01-05&0.058&0.264&4.89&27.71&10.96&12.03\\
\hline 
1956-01-06&0.057&0.282&4.69&28.84&11.14&11.04\\
\hline 
1956-01-08&0.057&0.273&7.13&27.35&12.71&11.55\\
\hline 
1956-01-09&0.078&0.260&8.87&27.01&12.58&12.42\\
\hline 
1956-01-10&0.068&0.266&7.10&27.65&12.18&12.64\\
\hline 
1956-05-02&0.069&0.265&7.78&26.46&11.96&9.03\\
\hline 
1956-05-05&0.087&0.294&10.65&29.72&12.92&10.43\\
\hline 
1956-04-03&0.078&0.075&7.45&6.20&11.74&9.56\\
\hline 
1959-01-29&0.096&0.091&14.42&13.10&15.09&14.40\\
\hline 
1959-03-29&0.090&0.094&13.00&12.18&14.93&13.03\\
\hline    
  
\end{tabular}}
\end{center}
\end{table}

\subsection{Discussions and conclusion}

With the application of the ECT to the KO and MWO Ca-K images, the detection of small-scale features has become accurate (panel (b) of Figures~\ref{fig:4} – \ref{fig:6}). Most of the images were obtained in the Ca-K line center. But, some changes were made in the instruments over the long period of observations over a century at KO and MWO. Therefore, a significant portion of the data has (1) high contrast images (larger FWHM of the intensity distribution of the image) due to the small passband of the instrument and/or setting of exit-slit of spectroheliograph at the blue-wing of the Ca-K line and/or use of high contrast emulsion, etc.; (2) low contrast images (smaller FWHM of the intensity distribution of the image) because of large passband and/or setting of the exit slit away from Ca-K and /or use of low contrast emulsion and/or weather conditions (cloudy sky), etc. The ECT procedure has been applied to make all the images of uniform nature \citep{singh2021} by rescaling the intensity of images such that the FWHM of the intensity distribution lies between 0.10 -- 0.11. Earlier reports about the quiet network area have not made any distinction between EN, AN, and QN as we have done. Hence the total area of small-scale features (EN+AN+QN, hereafter called quiet network) needs to be compared with the earlier reported quiet network area. During the 19th solar cycle, MWO data was obtained with narrow passband \citep{tlatov2009} yielding large FWHM of the intensity distribution and high contrast. Table~\ref{tab:1} gives the FWHM of the intensity distribution before the application of the ECT for KO and MWO sample images during this period and values of the quiet network derived from these images before and after the ECT application. The top 12 rows indicate that FWHM and quiet network values for KO are very less as compared to those for MWO data. 

The quiet network area values for high contrast images obtained at MWO and low contrast images at KO on the same day before the ECT application indicate a significant difference between the two ($\sim$~20\%). The difference in quiet network areas appears to be correlated with the difference in FWHM values. The large difference is due to observational differences and needs to be corrected. The ECT application to the data makes the data of similar nature, and the difference in values decreases to $<$ 2\% only. The bottom three rows in Table~\ref{tab:1} show minor differences between FWHM of intensity distributions and quiet network areas for KO and MWO data before and after the ECT application to the images. After the ECT application, the derived quiet network area values may be considered the representative value for the Ca-K images obtained at the line center. The decrease or increase in the quiet network area may be regarded as the correction for the high and low contrast images, respectively.

\citet{foukal2001} have used 27 images only, 3 Ca-K images of the highest quality (not defined) for each minimum phase during 1914 -- 1996. They found that the quiet Sun network occupies between 10 and 19\% (see Figure 3 in their article) during the minimum phase of the solar cycle. The significant variations in their detected quiet network might be due to different contrast of the selected images. The images they selected might be of high contrast data yielding high values for quiet network areas, as we have found. Further, their data is very limited, and that too during the minimum period. 
Present uncorrected data (Figure~\ref{fig:3}, left in the current manuscript) shows that the quiet network varies between 5 -- 40 \% of the disk (aggregate of EN, AN, and QN) for MWO and KO data, considering all the data of about 77 years. The significant variations are due to the analyzed images' very low and high contrast. After applying ECT to the images to make data of uniform contrast, the detected quiet network varies between 4 to 18 \% only (Figure~\ref{fig:7}) for the MWO and KO data, considering data for minimum, intermediate and maximum phases of the solar cycle together. It may be noted that the quiet network area (EN+AN+QN) occupies 4 -- 7 \% of the solar surface during the minimum phase and about 13 -- 18 \% during the maximum phase of the Sun (Panel (b) of Figures~\ref{fig:4} to \ref{fig:6}).
 
 
\citet{foukal1991} have analyzed 110 magnetograms taken during 1979, 1981, and 1986 to study the ratio between quiet network area and the total photospheric area outside active regions. The 15-20\% variations in the ratio (see Figure 4 in their article) do not refer to a fraction of the solar surface but only a fraction of that surface, excluding active regions. Moreover, only latitudes between $\pm$30 degrees were considered. In contrast, we believe the area of the whole visible solar disk leads to lower values. \citet{dere1984} studied the line profiles of C IV obtained at the quiet region (a small region of 10 $\times$ 800 arcsec) during the active period of 1979 and reported that the network at 1,00,000 K covers 16\% of the quiet solar surface. \citet{foukal2009} studied plage index (which they stated also included AN and EN contribution) and found solar cycle variation. Given the more significant contribution of plages to the plage index, a priori, it is not clear if the quiet Sun network should also vary with the cycle as plages do. Thus, the previous studies that either included EN and AN contribution to plage index or used selected limited subsets of data are unable to undoubtedly demonstrate the cycle variation of EN, AN, and QN components. 

The present analysis allows separating these components and showing their cycle variation directly for the first time. The fact that all active sun components (AN, EN, QN, plage, and active region) show variations in phase with sunspot number may suggest that area that is free of the magnetic field should vary in antiphase with the sunspot cycle (or that the magnetic fields are more tightly packed together during maxima of sunspot activity as compared with periods of minima).
 
The slope (31.2 for KO and 31.8 for MWO) and intercept (-245 for KO and -238 for MWO) of linear regression of the scatter plot between the percentage of total small-scale features and sunspot numbers agree with each other, as indicated in Figure \ref{fig:7}. Similarly, the slope values (27.6 and 28.4) for the KO and MLSO (PSPT) filtergrams agree, but the intercept (-122 and -217) are different. The slope values for filtergram data are less compared to those for spectroheliograms. This might be because the sunspot numbers are different for the spectroheliograms and filtergrams observing periods. The maximum sunspot number during spectroheliogram observations (1905 – 1997) is around 450 compared to $\approx$350 during filtergram observations (1997 – 2015). The scatter plots for spectroheliogram indicate a tendency for \% EN area not to vary with sunspot number $>$ 250. This can cause a difference in slope values. This can also be due to the broader passband of the filtergrams compared to that for spectroheliograms and different spatial resolutions. More sets of filtergram data with different specifications observed during the same period need to be analyzed to determine the reason for the differences. 

Considering the average normalized intensity of network features and variation in their areas, on average 3\% variation in the Ca-K flux occurs due to small-scale elements during a solar cycle. It may be noted that the total area occupied by the Ca-K features, representing a small-scale magnetic element,  varies from 5 to 20\% on the solar disk between the minimum and maximum phase of the solar cycle when considered on a daily basis. There can be two reasons for this increase. One, the Ca-K network size becomes smaller at the maximum phase as found by \citet{singh1981, pevtsov2016}. They reported that cell size is $\approx$ 5\% smaller at the maximum phase than at the minimum phase. This can lead to the larger area being occupied by the network boundaries. 

Variations in size of the supergranulation were reported by several studies \citep[for review, see:][]{rincon2018} with somewhat contradictory results. Most studies \citep[e.g.,][]{singh1981} found an anti-correlation with the solar cycle. Some other studies found the opposite correlation \citep[e.g.,][]{wang1988, muenzer1989, meunier2008}. \citet{tlatov2012} found a cycle variation in size of the supergranulation with a small phase shift with respect to the sunspot cycle. The phase shift was about 1.5 years, with minima/maxima in the size of the supergranulation occurring after sunspot minima/maxima. \citet{pevtsov2016} analyzed the annual mean spatial scales detected through a periodogram analysis and found less than 5\% variations over a solar cycle$;$  comparable with the scatter in data. Another reason may be that the magnetic field becomes stronger at the boundaries of a network at the maximum phase compared to that at the minimum phase. This can increase the brightness of the network boundaries during the maximum phase. Therefore, more network areas are detected during the maximum phase. It may be noted that the boundaries of the network are not uniformly bright. The brightness of the portion of the boundary depends on the underlying magnetic field.      
 
\par In summary, we can say that the ECT methodology improves the uniformity of Ca-K images taken over a long time with different instruments to generate extended series to study long-term variations on the Sun. The use of ECT has enabled us to reliably study the systematic variations in the small-scale Ca-K features, thereby the magnetic elements. The results of this study will help to make realistic modeling of irradiance and understand the dynamics of the Sun.

\begin{acknowledgements}

We thank the numerous observers who have made observations, maintained the data at KO and MWO. We also thank the digitization teams at KO (Founder PI, Jagdev Singh) and MWO.
The high spatial resolution Ca-K data was downloaded from Stanford Helioseismology Archive - prog$:$mwo (http$:$//sha.stanford.edu/mwo/cak.html). The Precision Solar Photometric Telescope was maintained and operated at Mauna Loa Solar Observatory from 1998 to 2015 by HAO/NCAR. The data was processed  and is served to the community by the Laboratory for Atmospheric and Space Physics, University of Colorado, Boulder (http$:$//lasp.colorado.edu/pspt access/). The sunspot data is from the World Data Center SILSO, Royal Observatory of Belgium, Brussels. The National Solar Observatory (NSO) is operated by the Association of Universities for Research in Astronomy (AURA), Inc., under a cooperative agreement with the National Science Foundation. Observations at the Mount Wilson Observatory have been supported over the years by the Carnegie Institute of Washington, the National Aeronautics and Space Agency, the US National Science Foundation, and the US Office of Naval Research. L.B. and A.A.P. are members of the international team on Modeling Space Weather And Total Solar Irradiance Over The Past Century supported by the International Space Science Institute(ISSI), Bern, Switzerland and ISSI-Beijing, China.
\end{acknowledgements}

\bibliographystyle{raa}
\bibliography{raabib}

\end{document}